\documentclass[sigconf,nonacm,natbib=false]{acmart}

\usepackage{booktabs}       
\usepackage{amsfonts}       
\usepackage{nicefrac}       
\usepackage{microtype}      
\usepackage{lipsum}
\usepackage{adjustbox}
\usepackage{graphicx}
\usepackage{caption}
\usepackage[misc]{ifsym}
\usepackage[super]{nth}
\usepackage{subfigure}
\usepackage{amsmath,amsfonts,amsthm,bm}
\usepackage{multirow}
\usepackage{array}
\usepackage{mathtools}
\usepackage{bbm}
\usepackage[normalem]{ulem}
\usepackage{wrapfig}
\usepackage{soul}
\usepackage{stfloats}
\usepackage{algorithm}
\usepackage{algorithmic}
\usepackage{setspace}
\usepackage{wrapfig}
\usepackage{enumitem}
\usepackage{caption}
\usepackage{makecell}
\usepackage{subfigure}

\setlist{topsep=0pt, leftmargin=*}
\usepackage[style=numeric,sorting=none]{biblatex}  

\AtBeginDocument{%
  \providecommand\BibTeX{{%
    \normalfont B\kern-0.5em{\scshape i\kern-0.25em b}\kern-0.8em\TeX}}}

\copyrightyear{}
\acmYear{}
\acmDOI{}

\thanks{\textbf{Preprint Version}: This is the preprint version of the paper.}

\begin{document}
\sloppy

\title{Uni-QSAR: an Auto-ML Tool for Molecular Property Prediction}

%
\author{Zhifeng Gao}
\authornote{Theses authors contributed equally to this research.}
\orcid{0000-0001-8433-999X}
\affiliation{%
  \institution{DP Technology}
  \streetaddress{No. 2, Haidian East 3rd Street}
  \city{Beijing}
  \country{China}
  \postcode{100080}
}
\email{gaozf@dp.tech}

\author{Xiaohong Ji}
\authornotemark[1]
\affiliation{%
  \institution{DP Technology}
  \streetaddress{No. 2, Haidian East 3rd Street}
  \city{Beijing}
  \country{China}}
\email{jixh@dp.tech}

\author{Guojiang Zhao}
\authornotemark[1]
\authornote{This work was conducted during these authors' internship at DP Technology.}
\affiliation{%
  \institution{Carnegie Mellon University}
  \streetaddress{5000 Forbes Ave}
  \city{Pittsburgh}
  \country{United States}}
\email{guojianz@andrew.cmu.edu}

\author{Hongshuai Wang}
\orcid{0000-0003-3353-372X}
\authornotemark[2]
\affiliation{%
  \institution{Soochow University}
  \streetaddress{No. 199, Renai Road}
  \city{Suzhou}
  \country{China}}
\email{hswang@stu.suda.edu.cn}

\author{Hang Zheng}
\affiliation{%
  \institution{DP Technology}
  \streetaddress{No. 2, Haidian East 3rd Street}
  \city{Beijing}
  \country{China}}
\email{zhengh@dp.tech}

\author{Guolin Ke}
\authornote{Guolin Ke and Linfeng Zhang are the corresponding authors.}
\affiliation{%
  \institution{DP Technology}
  \streetaddress{No. 2, Haidian East 3rd Street}
  \city{Beijing}
  \country{China}}
\email{kegl@dp.tech}

\author{Linfeng Zhang}
\authornotemark[3]
\affiliation{%
  \institution{DP Technology}
  \streetaddress{No. 2, Haidian East 3rd Street}
  \city{Beijing}
  \country{China}}
\email{zhanglf@dp.tech}

%
\renewcommand{\shortauthors}{Gao, et al.}

\begin{abstract}
Recently deep learning based quantitative structure-activity relationship (QSAR) models has shown surpassing performance than traditional methods for property prediction tasks in drug discovery. 
However, most DL based QSAR models are restricted to limited labeled data to achieve better performance, and also are sensitive to model scale and hyper-parameters. 
In this paper, we propose Uni-QSAR, a powerful Auto-ML tool for molecule property prediction tasks. Uni-QSAR combines molecular representation learning (MRL) of 1D sequential tokens, 2D topology graphs, and 3D conformers with pretraining models to leverage rich representation from large-scale unlabeled data. 
Without any manual fine-tuning or model selection, Uni-QSAR outperforms SOTA in 21/22 tasks of the Therapeutic Data Commons (TDC) benchmark under designed parallel workflow, with an average performance improvement of 6.09\%. 
Furthermore, we demonstrate the practical usefulness of Uni-QSAR in drug discovery domains.
\end{abstract}

\ccsdesc[500]{Applied computing~Life and medical sciences}
\ccsdesc[500]{Computing methodologies~Machine learning}

\keywords{Drug Discovery, Quantitative Structure-Activity Relationship, Molecular Representation Learning, Deep Learning}



\maketitle
\section{Introduction}
Building quantitative structure-activity relationship (QSAR)~\cite{tropsha2010best} models based on chemical compounds with measured properties to predict the properties for another batch of chemical compounds is a widely used method in drug discovery~\cite{neves2018qsar}, environmental toxicology~\cite{lavado2022qsar}, and chemical risk assessment~\cite{kar2018impact}.
How to build a QSAR model with good prediction ability, taking molecular structure as input and molecular property as output is the key problem.

Recently, several deep learning based QSAR tools have been developed and widely used~\cite{intro6} in real-world applications, such as DeepAutoQSAR~\cite{dixon2016autoqsar}, ChemProp~\cite{chemprop1}, and DeepTox~\cite{deeptox} etc. 
These tools can be used to predict the activity of compounds for a variety of biological targets, including enzymes, receptors, and transcription factors ~\cite{intro1_9}. 
DeepAutoQSAR is based on AutoML, which ensembles a collection of different models and hyper-parameters by ranking their fitness. 
Chemprop is another popular QSAR tool by ensembling message passing neural networks with different data splits, which is shown to be effective in predicting the activity of compounds for a range of targets, including enzymes and receptors. 
DeepTox, built from CNN networks, has been used to predict the toxicity of chemicals for a variety of endpoints, including skin irritation and genotoxicity. \looseness=-1

However, the labeled QSAR data from experiments are often scarce due to the high experimental costs. Those supervised learning based QSAR tools usually cannot perform well on small labeled data, suffering from over-fitting and difficulty in extrapolating.
Fortunately, pre-training (or self-supervised learning) is a recent effective technology to improve the performance in learning from small labeled data ~\cite{yu2021review}. 
To further improve the effectiveness of QSAR, we can leverage the pretrained molecular representation learning (MRL) models.
Earlier MRL approaches, such as molecular fingerprints and descriptors, are carefully designed by artificial experts to represent the properties of molecular parts. They are widely used, but the prediction ability of QSAR models based on them is relatively limited~\cite{intro1, intro2, intro3}. 
Along with the spurt of deep learning and the self-supervised learning (pretraining) technology ~\cite{intro13, molclr, self_sup_1, self_sup_2, self_sup_3}, pretrained MRL models are becoming prevalent. In particular, we can categorize the pretrained MRL methods into the following 3 classes.
\begin{itemize}
    \item \emph{1D MRL (1D Sequential Tokens)}. String based molecular structure representations have been developed since the 1960s, and SMILES~\cite{2dgraph1}, InChi~\cite{heller2013inchi} are the most popular ones. With the success of Transformer based pretrained methods in natural language areas, several studies~\cite{intro7,intro8,intro9} use BERT-like architectures with random masked SMILES pretraining tasks on large-scale unlabeled SMILES strings, and these methods have achieved some good results on downstream tasks than supervised learning.
    \item \emph{2D MRL (2D Topology Graphs)}. Graph based molecular structure representations are widely used in pharmaceutical and chemical staffs due to more explicit and abundant information than SMILES. Molecular graph can be built with chemical atoms, groups and bonds. With many graph neural networks based MRL proposed~\cite{intro10} recently, researchers developed various pretraining strategies, including atoms or bonds masking prediction~\cite{intro11}, 2D-3D transfer learning~\cite{stark20223d, liu2021pre}, contrastive learning~\cite{molclr} etc. The shared parameters in these GNN-based MRL methods are usually much smaller than those in pre-training methods based on transformers, restricting the exploitation of molecular information. It's worth noting that 2D-3D transfer learning methods often use 2D graphs as input for downstream tasks, but utilize 3D information during pre-training, thus categorizing these methods as 2D MRL.
    \item \emph{3D MRL (3D Conformers)}. The 3D molecular conformation is the natural representation in physical space that indicates a set of atoms with 3D spatial positions. In pure 3D pretraining MRL, researchers want to directly learn information from conformation sampling space, bridging the gap of many 3D related downstream tasks with pretraining. Uni-Mol~\cite{intro18} is the first pure 3D pretraining MRL that directly recovers coordinates and predicts masked atoms in pretraining with a SE(3) transformer architecture, and it is proven in a variety of downstream tasks, especially structure related tasks.
\end{itemize}

\vspace{5pt}

Among them, the 3D MRL models perform the best, since the properties of molecules are mostly determined by their 3D structures~\cite{crum1865connection,hansch1964p}, and Uni-Mol\cite{intro18}, a pure 3D MRL model, explicitly proven it in a variety of downstream tasks. However, there are several tasks that are less sensitive to 3D Conformers rather than some key functional groups like hERG~\cite{sanguinetti2006herg}, and local chemical environments like pKa prediction~\cite{perrin1981pka}. Besides, unlike the easy-obtained 3D conformers used in pretrained, the bioactive 3D conformers (such as the ligand conformers in the protein-ligand complexes) are hard to obtain but necessary for some downstream tasks, like bio-activity prediction tasks. In short, the 3D MRL is not always the best in various QSAR tasks~\cite{dare2022conformation,guimaraes2016conformation}.

\graphicspath{{}}
\begin{figure*}
    \centering
    \includegraphics[width=6.5in]{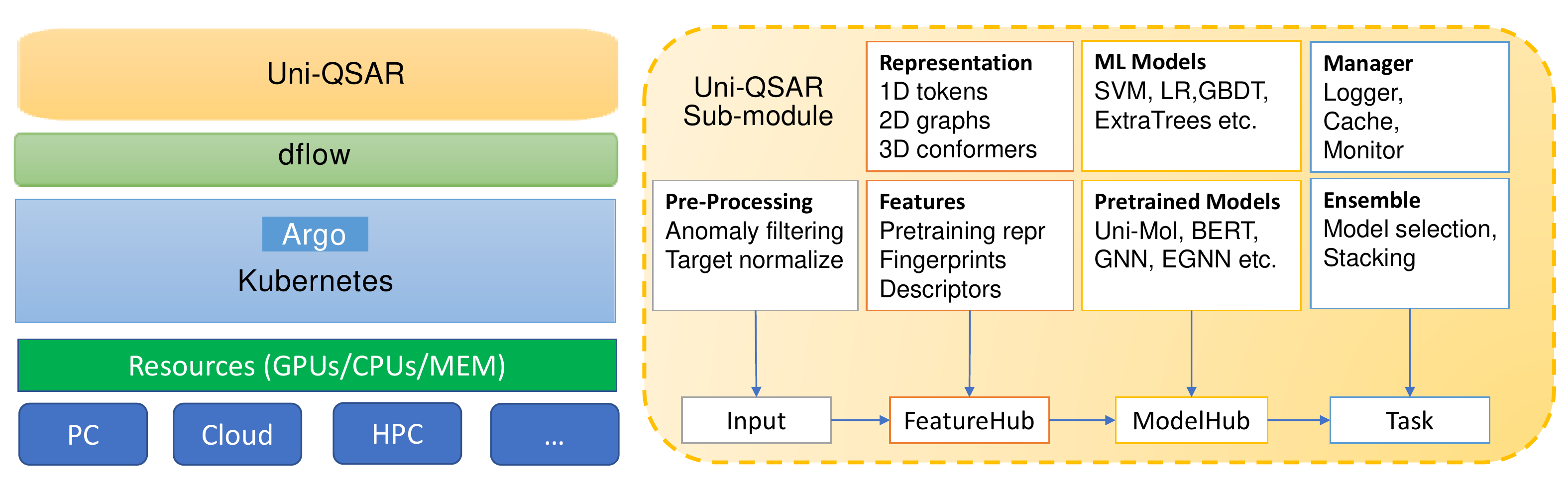}
    \caption{Schematic illustration of Uni-QSAR, left figure indices to the parallel framework of QSAR workflow, the right figure provides important sub-modules of Uni-QSAR, including Input Processing, FeatureHub, ModelHub, and Task, which constitutes jobs to run in the workflow.
    }
    \label{fig:uniqsar-workflow}
\end{figure*} 

To address that, we propose Uni-QSAR, which combines multiple MRL models. And to our best knowledge, Uni-QSAR is the first QSAR tool that combines MRL models of 1D sequential tokens, 2D topology graphs, and 3D conformers with large-scale unlabeled data pretraining models. Moreover, Uni-QSAR is an Auto-ML tool without any manual fine-tuning or human intervention in hyper-parameters and model selections under an efficient parallel workflow, bringing expediency to professional pharmaceutical and chemical scientists in many drug discovery areas, including 
virtual screening of chemical libraries~\cite{shoichet2004virtual}, in silico ADME/T properties prediction~\cite{dong2018admetlab}, Deep Docking~\cite{gentile2020deep} etc. 
Additionally, our framework aims to be efficient and automated, using the Dflow \cite{dflow} framework to build scientific computing workflows. This enables Uni-QSAR to make full use of the computing resources with minimal effort. An overview of our framework and sub-modules is shown in Fig.~\ref{fig:uniqsar-workflow}

We conduct extensive experiments on ADME/T property prediction tasks in TDC~\cite{huang2021therapeutics} benchmark comparing with other popular QSAR tools. Results demonstrate that Uni-QSAR achieves 6.09\% in average better performance. Furthermore, Uni-QSAR is successfully applied to construct chemical molecular libraries for CNS drugs, with a beneficiation process from over 1 billion chemical database to 1.3 million. Ablation studies show that a pure 3D pretraining model significantly improves QSAR tool's average performance, and the designed auto ml strategies, including target normalization, auto stacking etc, solidly contribute to the final performance. 

Our contributions are summarized as follows:
\begin{itemize}
    \item We introduce the first pretrained MRL based QSAR framework named Uni-QSAR, which combines 1D tokens, 2D graphs and 3D conformers with large-scale pretraining methodology, and we show 3D MRL plays an important role in molecular property prediction.
    \item We propose an Auto-ML and easy-to-use QSAR tool with multi-platform support and dynamic resource pool allocation, bringing expediency and efficiency to the pharmaceutical and chemical community.
    \item Experimental results show that our tools outperform SOTA in various molecular property prediction tasks, especially in ADME/T, which can solidly accelerate the process in drug discovery areas. 
\end{itemize}

\section{Method}
\subsection{Molecular Representation Learning}
Molecular representation learning aims to encode molecules into numerical vectors using deep learning or machine learning models, thereby preserving useful molecular information and serving as feature vectors for downstream tasks. Recently, many methods have been specifically designed for molecular representation learning, which can better integrate knowledge from the chemical domain. However, these methods usually require labeled molecular data to guide supervised training, but labeled data is usually scarce, and the acquisition of labeled molecular data is time-consuming and expensive, therefore, methods using machine learning and deep learning often lead to challenges of overfitting and poor generalization \cite{intro10, wu2018moleculenet}.


To address the problem of insufficient labeled molecules while hoping to simultaneously learn rich molecular structure and semantic information from large amounts of unlabeled molecular data, an increasing number of pre-training methods are using unlabeled molecular data in order to gain more information about the molecules, which leads to better generalization ability of the model. Additionally, using molecular fingerprints and descriptors as prior information can improve the performance of the model to some extent.

\begin{figure}[h]
    \centering
    \includegraphics[width=8.cm,height=4cm]{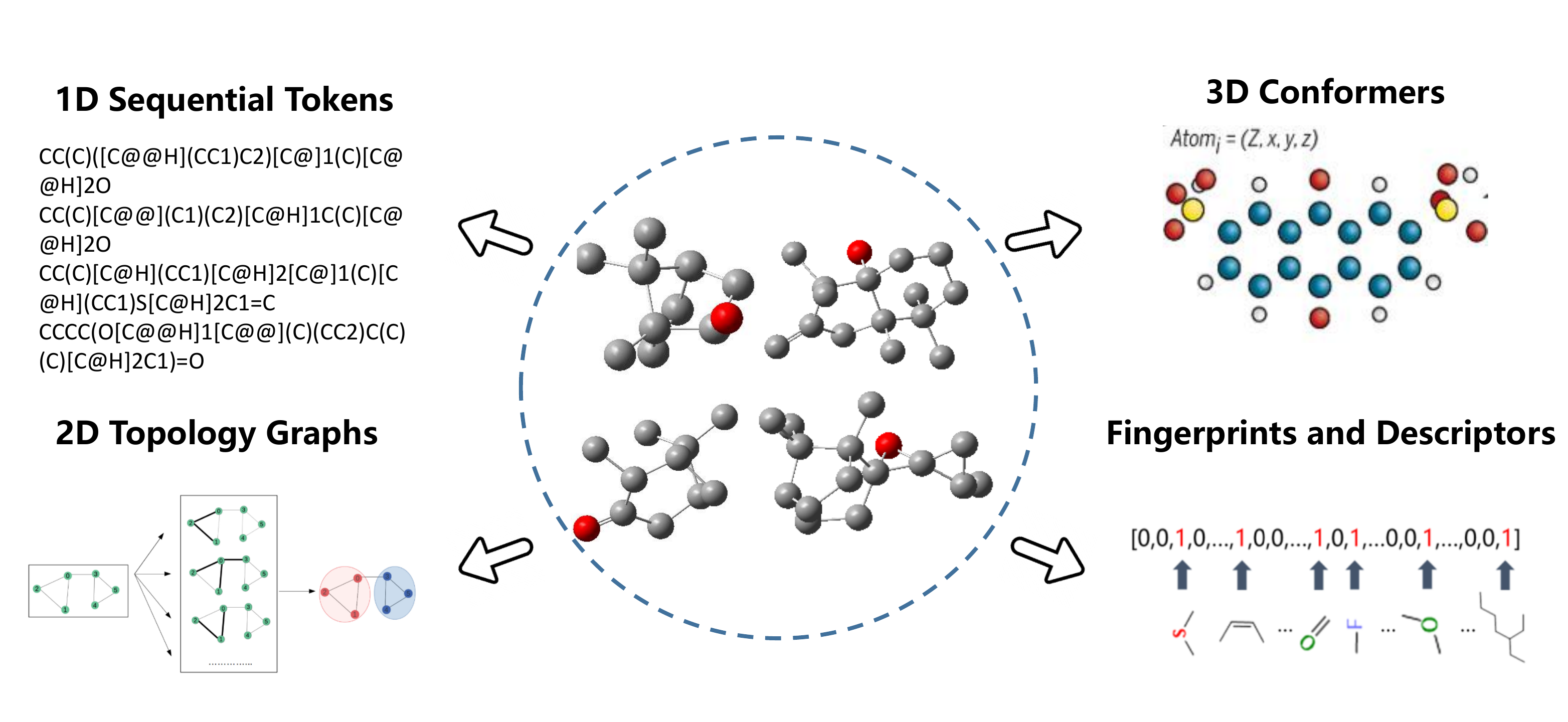}
    \vspace{-10pt}
    \caption{Different types of molecular representations used in Uni-QSAR. 
    1D: SMILES
strings that use simplified text encodings to describe the structure of a chemical species. 2D: Graph with atom and bond. 3D: Atomic geometric position coordinates. Fingerprint vector: Quantifies presence or absence of molecular environments}
    \label{fig:mrl}
\end{figure}

To acquire more information about the molecules in order to improve the generalization ability of the model for downstream tasks, Uni-QSAR uses a mixture of traditional fingerprints and descriptors
 \cite{fp} with 1D sequential tokens \cite{intro8}, 2D topology graphs \cite{intro13, molclr, hignn, kpgt} and 3D conformers \cite{intro18, EquivariantNN} under pretraining models. As shown in Fig.~\ref{fig:mrl} of all those molecular representations, more details will be discussed next.

\graphicspath{{}}

\paragraph{Fingerprints and Descriptors}
In addition to neural network models, some classical machine learning algorithms such as Extreme Tree, Gradient Boosting Decision Tree, Support Vector Machine, etc. are also integrated into our framework. Due to the weak ability of these models to directly extract chemical structure information, some molecular fingerprints and descriptors with prior information are added. 
Molecular fingerprints are typically very high-dimensional and are often used in conjunction with a similarity measure such as the Tanimoto coefficient \cite{tanimoto} to compare the similarity of different molecules. In our framework, we use Morgan fingerprints \cite{morgan}, a type of molecular fingerprint that encodes the structural and chemical properties of a molecule into a fixed-length numerical representation. 
Molecular descriptors are continuous numerical representations of a molecule. These representations are often lower-dimensional than fingerprints and can capture more detailed information about the structure and properties of a molecule\cite{moldesc}. We use many different types of molecular descriptors in our framework including:
\begin{itemize}
\item Structural descriptors: These describe the structure of a molecule, and can include measures such as the number of atoms, bonds, and functional groups, as well as more complex measures such as topological indices and graph invariants.

\item Topological descriptors: These are numerical values that describe the topological features of a molecule, such as the number of atoms, bonds, and rings. Examples include the BertzCT index and the Balaban index.

\item Constitutional descriptors: These describe the arrangement of atoms and bonds within a molecule. Examples include the Randic index and the Zagreb index.

\item Physicochemical indices: 
These describe the physical and chemical properties of a molecule, such as 	The average molecular weight of the molecule and the polar surface area of a molecule based upon fragments.

\item Connectivity descriptors: Connectivity descriptors are a type of molecular descriptor that is based on the connectivity of atoms within a molecule. These descriptors describe the arrangement of atoms and bonds in a molecule. Examples include the Chi indices.
\end{itemize}

\paragraph{1D Sequential Tokens}
Transformers ~\cite{1d1,1d2}, are gradually becoming the de-facto standard for string-based self-supervised representation of language models. Inspired by this trend, several recent works have  used transformers for molecular property prediction and have obtained promising results \cite{1d3,1d4}. Considering the availability of billions of SMILES strings, Transformer offers an interesting alternative to expert fingerprints. Some studies  \cite{intro7,intro8,1d5} have used the BERT-like transformer models to design pre-training tasks such as atomic prediction tasks, molecular feature prediction tasks, and contrastive learning tasks for SMILES on large-scale unlabeled data, and have achieved good results on downstream tasks. Our framework is capable of adapting pre-training models to leverage sequential SMILES tokens for downstream tasks. As an example, we use a Knowledge-based BERT (K-BERT) \cite{intro8} pre-training model that enhances the ability to extract SMILES molecular features through  well-designed pre-training strategies. 

\paragraph{2D Topology Graphs} 
In order to overcome the limitations of previous string-based representations  of molecules, which were unable to encode important topological information \cite{2dgraph1}, researchers have turned to using graph neural networks  \cite{2dgraph2, 2dgraph3} in their studies.
  However, graph neural networks have the following problems in the application of molecules: 1) Insufficient labeled molecules for training; 2) Poor generalization ability to newly synthesized molecules \cite{intro13}. Our pipeline is compatible with the following different paradigms to solve the above problems.

\begin{itemize}
\item 
Leverages node- and edge-level contextual property prediction tasks, as well as self-supervised graph-level motif prediction tasks to learn the rich molecular structure and semantic information from a large quantity of unlabeled molecular data.
\item
Introduces a method for learning molecular representations that involve contrasting positive and negative pairs of molecular graphs and employs atom masking, bond deletion, and subgraph removal enhancement strategies. 
\item
Proposes knowledge-oriented pre-training strategy with additional knowledge of molecules to guide the prediction of randomly masked nodes. 
\item
Presents a hierarchical information graph neural network framework that is designed to predict the characteristics of molecules by using both molecular graphs and fragment information. 
\end{itemize}
Our framework can easily adapt 2D topology graph pre-trained models on downstream tasks. Due to space constraints, here we just mention a few of the models we use according to different paradigms, which are GROVER \cite{intro13} , MOLCLR \cite{molclr} , KPGT \cite{kpgt} , and HIGNN \cite{hignn}.

\paragraph{3D Conformers} 
Most methods in molecular representation learning (MRL) treat molecules as one-dimensional sequential tokens or two-dimensional topological graphs, which limits their ability to incorporate three-dimensional (3D) information for downstream tasks, such as 3D geometry prediction or generation. This is a significant drawback, as the properties of molecules and the effects of drugs are largely determined by their 3D structures in the field of life sciences \cite{3d14, 3d15}. As a result, the ability to effectively incorporate 3D information in MRL methods is crucial for accurately predicting or generating molecular structures. 

There have been recent efforts to incorporate 3D information in molecular representation learning (MRL) \cite{3d1,3d2}, but the performance of these methods has not been satisfactory. This may be due to the limited size of available 3D datasets and the fact that 3D positions cannot be used as inputs or outputs during fine-tuning, but only as auxiliary information. These limitations have hindered the effectiveness of incorporating 3D information in MRL.

To alleviate the above challenges, EGNN \cite{egnn} can take 3D information as input or output, which is essential in 3D spatial tasks. Our pipeline also provides high-level neural network layers for constructing SOTA SE3-equivariant neural networks based on Equivariant-NN-Zoo \cite{EquivariantNN}.

Moreover, Uni-Mol \cite{intro18} is the first simple and effective universal 3D molecular pretraining framework that is derived from a large amount of unlabeled data. It is able to take 3D positions as both input and output and utilizes a 3D position denoising task to learn 3D spatial representations with SE(3)-equivariant transformer in addition to masking atomic predictions in the pretraining tasks. Our framework has demonstrated promising performance on various downstream tasks through the utilization of Uni-Mol's pre-trained model.

\subsection{Auto strategies}
As an Auto-ML tool, Uni-QSAR integrates several empirical and theoretical strategies together to achieve good performance. In pre-processing, anomaly detection with 3-sigma rule is applied. During training, auto target normalization for regression tasks and loss optimization for imbalance tasks is very crucial. For model ensembling, Auto stacking is used to get better and more stable performance. More details are attached below.

\paragraph{Target Normalization} 
Models benefit from the normalization of targets, notably, in many molecule property regression tasks, target is extremely skewed due to inconsistent experimental and observational methods. Previous methods usually use standardization to ignore the shape of the distribution by removing the mean and deviation of target. In Uni-QSAR, auto target normalization with a combination of non-linear power transformation~\cite{carroll1981prediction} and standardization transformation is applied, Box-Cox~\cite{box1964analysis} and Yeo-Johnson~\cite{weisberg2001yeo} are used to transform original target distribution to nearly Gaussian distribution, and the specific algorithm is described in Alg.~\ref{alg:targetnormal}.
\begin{algorithm}[th]
\caption{Auto Target Normalization for Regression.} \label{alg:targetnormal}
\begin{algorithmic}[1]
\REQUIRE Raw training target $Y_{i}$ for $class_{i}$: $\mathbf{Y_{i}} \in \mathbb{R}^{m} $, training size $m$, skew threshold $v$
\IF{$\mathbf{\mathbf{skewness(Y_{i})}} <= \mathbf{v} $}
\STATE $\mathbf{\hat{Y}_{i}} = \mathbf{StandardTransform(Y_{i})}$ \COMMENT{standard Gaussian transform}
\ELSE
\IF {$\min(Y_{i})>0$} 
\STATE $\mathbf{PowerTransform} = Box Cox$
\ELSE \STATE $\mathbf{PowerTransform} = YeoJohnson$
\ENDIF
\STATE $\mathbf{\hat{Y}_{i}} = \mathbf{PowerTransform(Y_{i}, \lambda \in \mathbb{R})}$ \COMMENT{power transform fitting for best $\lambda$}
\ENDIF
\RETURN $\mathbf{\hat{Y}_{i}}$
\end{algorithmic}
\end{algorithm}

\paragraph{Loss Optimization}
In QSAR modeling, datasets are usually highly imbalanced, contain very few functional candidates or interested classes, and hundreds or thousands of times more unfunctional and uninterested molecules~\cite{guan2022class}. The minority interested class is ill-sampling and biased which can lead to poor generalization performance. In this work, we investigate several loss functions, including Focal loss~\cite{lin2017focal} and GHM loss~\cite{li2019gradient} for imbalance tasks. Focal loss tries to re-weight positive and negative classes concurrently with amplifying hard learning examples by two hyper-parameters $\alpha$ and $\gamma$, formally as:
\begin{equation}
FL(p_{t}) = -\alpha_{t} (1-p_{t})^\gamma\log(p_{t})
\end{equation}
However, in many molecule properties prediction tasks, especially in bioactive and drug discovery areas, the interested class is more likely to be predicted as a rare occurrence, ignored altogether, or assumed as noise or an outlier~\cite{guan2022class}. In these imbalance tasks, hard examples contain a high range of outliers. Gradient Harmonizing Mechanism(GHM) tries to solve this by using Gradient Density to balance the gradient flow with hard and easy examples, formally as:
\begin{equation}
GD(g) = \frac{1}{l_{\epsilon}(g)}\sum_{k=1}^{N} \delta_{\epsilon}(g_k,g)
\end{equation}
\begin{equation}
GHM(p_{t}) = \frac{CE(p_{t},p)}{GD(g_t)}
\end{equation}
In Uni-QSAR, we first detect imbalance tasks with a simple threshold rule of class proportion calculated by the training set, then applied focal loss and GHM loss independently and select under CV performance in the parallel workflow.
Details of Focal loss and GHM loss can be seen in the origin paper, hyper-parameters setting will be shown in the appendix.

\paragraph{Auto Stacking}
\graphicspath{{}}
\begin{figure*}[t]
    \centering
    \includegraphics[width=6.8in,height=6.5cm]{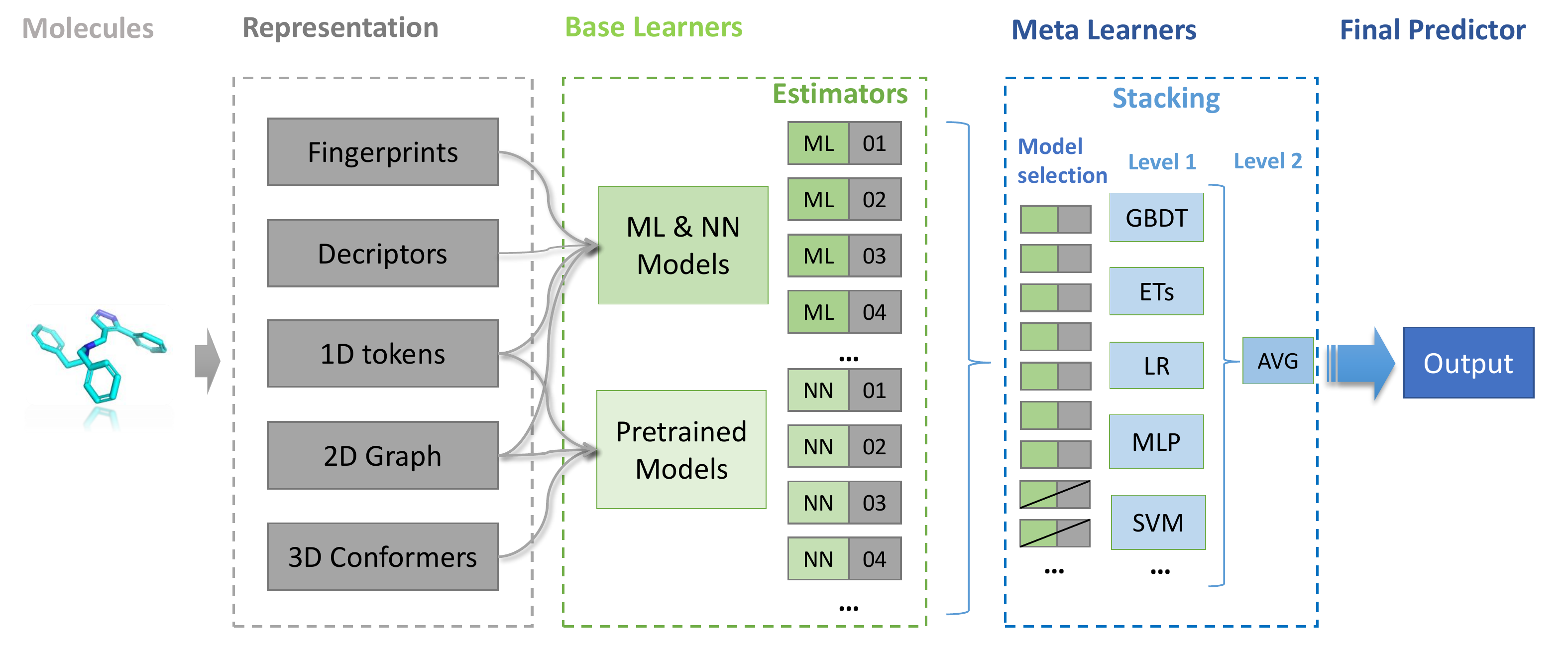}
    \vspace{-10pt}
    \caption{Uni-QSAR Auto Stacking Overview: Base Learners indices search space of estimator, including different combination of features and models. Meta Learners take estimator's output as stacking model's input, then several ML models are applied to fitting this. All processors are in automatic schema, including model selection and parameter tuning.}
    \label{fig:uniqsar-stacking}
\end{figure*} 

In ensemble learning, bagging~\cite{breiman1996bagging} and boosting~\cite{zhou2012ensemble} are widely used due to their simplicity. Stacking~\cite{wolpert1992stacked} directly uses the base learners as the input features to improve the predictive ability of the classifier which can enlarge the ensembling power of different models. Also stacking is widely used in winner teams' solutions for data mining competitions. In QSAR models, accuracy and stability are very curial, and the combination of molecular representation with model architecture is in varied forms. So we design a two-level stacking framework to ensemble meta-models of 1D, 2D, and 3D representations. As shown in Fig.~\ref{fig:uniqsar-stacking}, we combine several molecular representations as input of various base learners automatically, including machine learning, neural networks and pretraining models. Model selection is applied to filter meta estimators with lower cross-validation scores, remaining estimators are trained using k-fold splits with the same seeds to eliminate over-fitting, it is noteworthy that we use a two-level stacking framework with a simple average operation added in level two to enhance stacking learning ability, which boosts the final performance with limited cost.

\subsection{Uni-QSAR Workflow} 
\begin{figure}[t]
    \centering
    \includegraphics[width=3.2in]{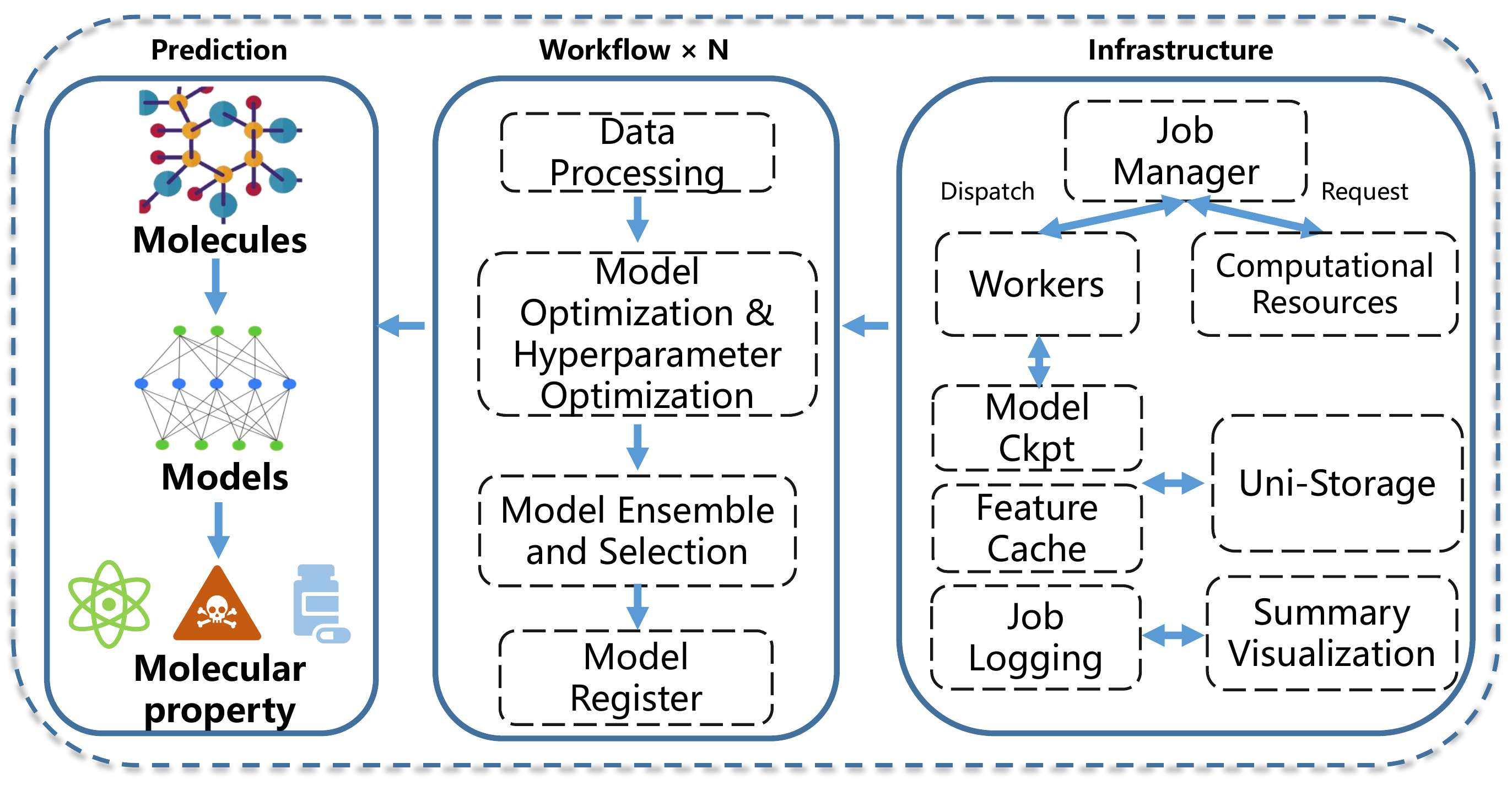}
    \caption{Uni-QSAR workflow Overview: The general workflow for a single job, the infrastructure behind the job execution and the inference pipeline using the learned model with Uni-QSAR.}
    \label{fig:uniqsar-workflow-inf}
\end{figure}

Uni-QSAR employs ensembles of machine-learning models to predict molecular property prediction. It will fit a collection of models with different architectures and hyper-parameters, training each model with a different strategy. Uni-QSAR achieves this in an automated and efficient fashion by building a workflow on Dflow \cite{dflow}. Dflow is designed to be based on a distributed, heterogeneous infrastructure, it's able to work on various resources. By developing a workflow based on Dflow, we can easily dispatch each model to different machines with scalability. Fig. \ref{fig:uniqsar-workflow-inf} gives an overview of the Uni-QSAR workflow, comprising three parts: 1) the general workflow for a single job; 2) the infrastructure behind the job execution. 3) The inference pipeline using the learned model with Uni-QSAR.

From the user's perspective, Uni-QSAR takes the smiles as input and performs model optimization and hyperparameter optimization to produce a domain-specific model for molecular property prediction. The ML modeling process of molecular property prediction was performed by Uni-QSAR automatically, the user did not need to care about the details of the implementation. 

Inside the block box ('workflow' part in Fig.\ref{fig:uniqsar-workflow-inf}), the input smiles will be preprocessed to fingerprints, descriptors, 1d tokens, 2d graphs, and 3d conformers. In the meantime, missing values filled and target normalization will be performed based on the skewness. Afterward, models were constructed and their parameters were optimized by performing Bayesian Optimization. Finally, a two-level stacking ensemble method was performed to enhance the final result.

From the implementation perspective ('infrastructures' part in Fig.\ref{fig:uniqsar-workflow-inf}), The foundation of Uni-QSAR is a parallel training workflow based on Dflow. The job manager can automatically request machines from the computational resources pool for workflow execution. Each model was trained on a different machine but shared the same precomputed feature pools. It achieved 2x speedup on our experiments and saved the idle GPU resource when training traditional machine learning models. After the workflow finished, the trained model was registered in Uni-storage for the model.

\graphicspath{{}}
\section{Experiment}
\subsection{ADMET properties prediction}
\paragraph{Datasets} It's wildly accepted that ADMET (absorption, distribution, metabolism, excretion, and toxicity) is the cornerstone of small molecule drug discovery, defining drug efficacy and toxicity profile. And The TDC \cite{huang2021therapeutics} ’s provided 66 AI-ready datasets spread across 22 learning tasks and spanning the discovery and development of safe and effective medicines. 
For our experiments, we run our benchmarks using the ADMET Benchmark Group consisting of 18 ADME and 4 Tox tasks with the same configuration as Schrodinger's benchmark report.
More details can be found in Appendix Table \ref{table:datasets}.

\paragraph{Baselines}  
Chemprop \cite{chemprop1,chemprop2}, is an open source Auto-ML based tool\footnote{https://github.com/chemprop/chemprop} for molecular property prediction. The core model of Chemprop is a directed message-passing neural network that iteratively aggregates local chemical features to predict properties. The Chemprop D-MPNN shows strong molecular property prediction capabilities across a range of properties, from quantum mechanical energy to human toxicity. Deep-AutoQSAR \cite{dixon2016autoqsar} was another automated machine-learning application to build, validate and deploy quantitative structure-activity relationship (QSAR) models tools built by Schrodinger \footnote{https://www.schrodinger.com/products/deepautoqsar}. Deep-AutoQSAR ensembles a collection of models with different architectures and hyper-parameters by ranking their fitness.
DeepPurpose \cite{deeppurpose} implements a unified encoder-decoder framework\footnote{https://github.com/kexinhuang12345/DeepPurpose} to rapidly build prototypes. By simply specifying the name of the encoder, the user can automatically connect the encoder of interest to the associated decoder. This library has achieved good results when used for predicting drug-target interactions, predicting compound properties, predicting protein-protein interactions, and predicting protein function.

\paragraph{Results}
Table \ref{table:regression}, \ref{table:classification} show the experiment results of our framework and competitive baselines. The ChemProp and Deep-AutoQSAR's results are from Schrodinger's white paper \cite{schrodinger_white_paper}, and the results of Deeppurpose come from the leaderboard of TDC benchmark\footnote{https://tdcommons.ai/benchmark/admet\_group/overview/}.  In addition, in order to illustrate the effect of our framework, we also add the ranking of our results on the leaderboard of different datasets. It's worth noting that our framework was able to achieve excellent results on multiple datasets with a single set of parameters. Overall, The Uni-QSAR outperforms Deep-AutoQSAR in all attribution prediction tasks with a 5.85\% improvement on average. Three key points lead to our method achieving so impressive results. First, Uni-QSAR integrates a pure 3D molecular representation \cite{intro18} model in our ensemble collections. Second, we use the auto-stacking method when selecting the bunch of collection models that contribute to the final results significant. Third, Uni-QSAR performs auto-target normalization and transformation when the target data is highly skewed. More details will be given in the ablation study section. Moreover, the Table \ref{table:regression_ab} results show that the performance will decline to varying degrees without using target normalization.


\begin{table*}
  \caption{Uni-QSAR performance on TDC ADMET Group benchmark(regression tasks)}
  \label{table:regression}
  \centering
   \addtolength{\tabcolsep}{-2pt}
   \begin{adjustbox}{max width=\linewidth}
  \begin{tabular}{l|lllll|llll}
    \toprule
    \multicolumn{1}{c}{} &\multicolumn{8}{c}{Regression}   &\multicolumn{1}{c}{}  \\
    \multicolumn{1}{c}{}  &\multicolumn{5}{c}{MAE (lower is better $\downarrow$)}  &\multicolumn{4}{c}{Spearman (higher is better $\uparrow$)}                 \\                  
    \midrule
    Datasets      & Caco2  & Lipo  & AqSol  & PPPR  & LD50  & VDss  & CL-Heap  & CL-Micro  & Half-Life \\
    \# Molecules  & 906 & 4200 & 9982 & 1797 & 7385 & 1130 & 1020 & 1102 & 667 \\
    \midrule
    ChemProp        & {0.390}     & {0.437}       & {0.820}    & {7.993}    & \textbf{0.548}      & {0.519}     & {0.423}  & {0.579}  & {0.293}  \\
    DeepAutoQSAR        & {0.306}     & {0.476}       & {0.784}    & {8.043}    & {0.59}      & {0.673}     & {0.432}  & {0.594}  & {0.551}  \\
    DeepPurpose        & {0.393}     & {0.574}       & {0.827}    & {9.994}    & {0.678}      & {0.561}     & {0.382}  & {0.586}  & {0.184}  \\
    \midrule
    Uni-QSAR        & \textbf{0.273}     & \textbf{0.420}       & \textbf{0.677}    & \textbf{7.530}    & {0.553}      & \textbf{0.729}     & \textbf{0.490}  & \textbf{0.645}  & \textbf{0.605}  \\
    TDC LB\textsubscript{Ranking} & \nth{1}/11 & \nth{1}/9 & \nth{1}/9 & \nth{1}/11 & \nth{1}/12 & \nth{1}/10 & \nth{2}/10 & \nth{1}/11 & \nth{1}/11 \\
    \bottomrule
  \end{tabular}
  \end{adjustbox}
\end{table*}

\begin{table*}
  \caption{Uni-QSAR performance on TDC ADMET Group benchmark(classification tasks)}
  \label{table:classification}
  \centering
   \begin{adjustbox}{max width=\linewidth}
  \begin{tabular}{l|llllllll|lllll}
    \toprule
    \multicolumn{1}{c}{} &\multicolumn{11}{c}{Classification}   &\multicolumn{1}{c}{}  \\
    \multicolumn{1}{c}{}  &\multicolumn{8}{c}{AUROC(higher is better  $\uparrow$)}  &\multicolumn{4}{c}{AUPRC(higher is better  $\uparrow$)}                 \\                  
    \midrule
    Datasets      & HIA  & Pgp  & Bioav  & BBB  & \thead{CYP3A4 \\ Substrate}  & hERG  & Ames  & DILI & \thead{CYP2C9 \\ Inhibition} & \thead{CYP2D6 \\ Inhibition} & \thead{CYP3A4 \\ Inhibition} & \thead{CYP2C9 \\ Substrate} & \thead{CYP2D6 \\ Substrate} \\
    \# Molecules & 578 & 1212 & 640 & 1975 & 667 & 648 & 7255 & 475 & 12092 & 13130 & 12328 & 666 & 664\\
    \midrule

    ChemProp        & {0.979}     & {0.902}       & {0.623}    & {0.882}    & {0.610}      & {0.750}     & {0.864}  & {0.918}  & {0.770} & {0.664} & {0.870} & {0.391}  &{0.688}\\

    DeepAutoQSAR        & {0.982}     & {0.917}       & {0.682}    & {0.876}    & {0.642}      & {0.845}     & {0.864}  & {0.933}  & {0.792} & {0.702} & {0.883} & {0.395}  & {0.703} \\
    DeepPurpose        & {0.972}     & {0.918}       & {0.672}    & {0.889}    & {0.639}      & {0.841}     & {0.823}  & {0.875}  & {0.742} & {0.616} & {0.829} & {0.380}  & {0.677} \\
    \midrule



    Uni-QSAR        & \textbf{0.992}     & \textbf{0.934}       & \textbf{0.732}    & \textbf{0.925}    & \textbf{0.645}      & \textbf{0.856}     & \textbf{0.876}  & \textbf{0.942}  & \textbf{0.801} & \textbf{0.743} & \textbf{0.888} & \textbf{0.454} & \textbf{0.721} \\
    TDC LB\textsubscript{Ranking} & \nth{1}/10 & \nth{2}/11 & \nth{3}/11 & \nth{1}/11 & \nth{3}/8 & \nth{3}/11 & \nth{1}/10 & \nth{1}/11 & \nth{1}/11 & \nth{1}/8 & \nth{1}/8 & \nth{1}/10 & \nth{1}/8 \\
    \bottomrule
  \end{tabular}
  \end{adjustbox}
\end{table*}

\subsection{CNS drugs enrichment}

\paragraph{Background} 
Due to the increasing aging population around the world, many aged people are suffered from CNS (central nervous system) diseases such as Alzheimer’s disease (AD), Parkinson’s disease (PD), brain cancer and etc, raise great demands for CNS drug development~\cite{ghose2012knowledge}. 
However, due to the limitation of relevant clinical animal models, it is difficult to test CNS drugs. And the lack of understanding of the complex mechanism of CNS pathogenesis, it is hard to simulate the process through physical modeling \cite{yu2022ensemble}. These challenges lead to a limited amount of labeled data related to CNS drugs.
Utilizing molecular representation learning (MRL) can achieve good results in data sets with small sample sizes such as CNS drugs.
By introducing the 3D pre-training model which is trained from large molecular data, Uni-QSAR would be more generalization compared with other QSAR methods.

\begin{figure*}[t]
\hspace{-1.0cm}
  \begin{minipage}[c]{0.75\textwidth}
     \footnotesize 
     \centering
     \includegraphics[width=11cm,height=4.0cm]{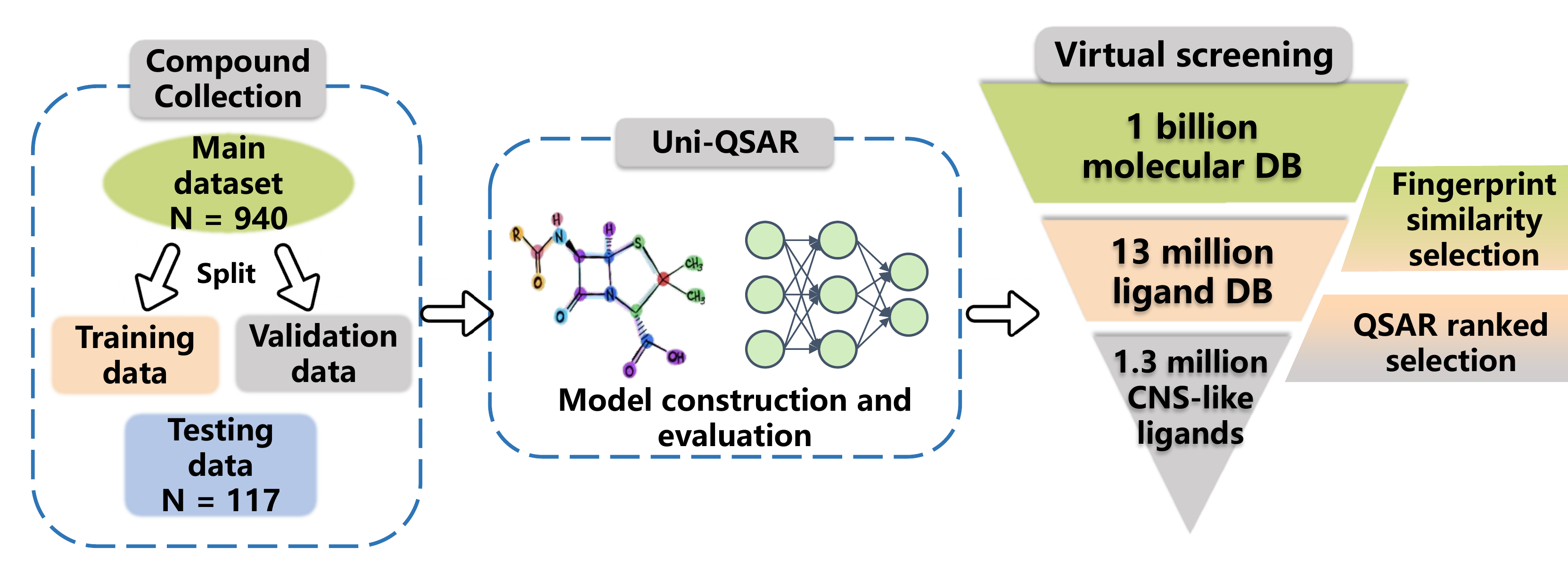}
     \vspace{2pt}
     \label{fig:uniqsar-virutal-screen-pipeline}
     \vspace{-4pt}
  \end{minipage}
  \hspace{-1.0cm}
  \begin{minipage}[c]{0.3\textwidth}
    \footnotesize 
    \centering
    \begin{tabular}{c|c|c}
    \toprule

    \multicolumn{3}{c}{Classification on CNS (random-split)} \\    
    \midrule
    Methods  & Validation AUC   & Testing AUC \\
    \midrule
    SVM & 0.948 & 0.958 \\
    GCN  & 0.971 & 0.873 \\
    Yu et al. \cite{yu2022ensemble} & 1.00 & 0.978 \\
    \midrule
    Uni-QSAR & 0.996 & \textbf{0.980} \\
    \bottomrule

    \end{tabular}
    \vspace{10pt}
    \hspace{10pt}
    \captionsetup{justification=centering}
    \label{cns-drug}
    \vspace{-4pt}
    \end{minipage}
    \caption{Uni-QSAR virtual screening pipeline for CNS drugs and Performance on CNS drug dataset. The left figure pipeline shows how to filter CNS-like ligands from the billions-level database, and the right table compared the performance of Uni-QSAR with other QSAR models on CNS drugs prediction task \cite{yu2022ensemble}}
    \label{cns drug all}
\end{figure*}

\paragraph{Data collection \& results} We build CNS models using Uni-QSAR with 940 market drugs \cite{ghose2012knowledge} (315 CNS-active and 625 CNS-inactive) as training data and evaluate models' performance using an additional external dataset of 117 market drugs \cite{van1998estimation} as testing data.
The result in Fig. \ref{cns drug all} shows that Uni-QSAR methods are more effective for the classification of CNS drugs versus non-CNS drugs. We compared a hybrid ensemble model 
proposed by Yu et al.\cite{yu2022ensemble}, which combines GCN and SVM with PaDEL descriptors 
 \cite{yap2011padel} and outperformed other constituent traditional QSAR models. However, our QSAR CNS model shows more generalization on the external testing data. 
 As shown in Fig. \ref{cns drug all} pipeline, our CNS drug enrichment or virtual screening process begins by utilizing fingerprints-based bit vectors for efficient similarity searching to coarsely screen 13 million molecules from the billions-level druggable database with Lipinski rule of five \cite{nogara2015virtual} and Veber rule 
\cite{pollastri2010overview} applied to filter no-druggable molecules. Then Uni-QSAR is used to finegrain selected 1.3 million molecules from above for the next phase of drug discovery processing, including lead optimization, ADMET prediction, and Re-Docking etc.


\section{Ablation study}
In this section, we conduct ablation experiments on molecular representation, auto stacking, and target normalization to analyze how each of the modules affects the performance of our framework. All experiments are based on the ADMET dataset.

\subsection{Molecular representation} 
As mentioned above, molecular representation learning (MRL) is a key step that automatically discovers the feature patterns in chemical science and the use of molecular 3D structural information can bring about effective improvements. Uni-Mol utilizes a 3D position denoising task to learn 3D spatial representations.
In order to explore the impact of the additional 3D structural information brought by Uni-Mol on our framework, we conducted an ablation study to uncover the role of molecular representation. Experiments in Table \ref{table:regression_ab} and \ref{table:classification_ab} show significant improvement by ensembling with the 3D molecular pre-training model, a.k.a Uni-Mol. 

\subsection{Auto stacking} 
Stacking allows for the combination of multiple models with good performance to produce better results than a single model. Uni-QSAR employs a two-level stacking framework to ensemble a collection of models. To gain a better understanding of the significance of using Auto-stacking in the framework, a comparison is done by averaging the predictions of various models. Experiments in Table \ref{table:regression_ab} and \ref{table:classification_ab} show that two-level stacking achieves improvement in almost all the tasks.


\subsection{Target normalization} 
Highly skewed data are common in regression tasks, it's harmful to model performance when target data is highly skewed. We conduct experiments to confirm that without target normalization Alg.\ref{alg:targetnormal}, most of the tasks with skewed data in regression, the performance is inferior to the model without using target normalization. The result is shown in Table.\ref{table:regression_ab}, combining with distribution analysis in appendix \ref{fig:distibution}, the task \textit{VDss} and \textit{Half Life} distribution with highly skewed before target normalization, the experiment result is not as good as which using target normalization.

\begin{table*}
  \caption{Uni-QSAR ablation study on molecular property prediction regression tasks}
  \label{table:regression_ab}
  \centering
   \begin{adjustbox}{max width=\linewidth}
  \begin{tabular}{l|lllll|llll}
    \toprule
    \multicolumn{1}{c}{} &\multicolumn{8}{c}{Regression}   &\multicolumn{1}{c}{}  \\
    \multicolumn{1}{c}{}  &\multicolumn{5}{c}{MAE (lower is better $\downarrow$)}  &\multicolumn{4}{c}{Spearman (higher is better $\uparrow$)}                 \\                  
    \midrule
    Datasets      & Caco2  & Lipo  & AqSol  & PPPR  & LD50  & VDss  & CL-Heap  & CL-Micro  & Half-Life \\
    \midrule


    Uni-QSAR w/o unimol       & {0.287}     & {0.506}       & {0.748}    & {7.964}    & {0.613}      & {0.716}     & {0.455}  & {0.630}  & {0.557}  \\

    Uni-QSAR w/o stacking       & {0.307}     & {0.542}       & {0.878}    & {8.151}    & {0.607}      & {0.703}     & {0.467}  & {0.60}  & {0.589}  \\
    


    Uni-QSAR w/o normalization       & {0.281}     & {0.423}       & {0.678}    & {7.598}    & {0.562}      & {0.660}     & {0.477}  & \textbf{0.657}  & {0.444}  \\
    
    \midrule

    Uni-QSAR        & \textbf{0.273}     & \textbf{0.420}       & \textbf{0.677}    & \textbf{7.530}    & \textbf{0.553}      & \textbf{0.729}     & \textbf{0.490}  & {0.645}  & \textbf{0.605}  \\
    \bottomrule
  \end{tabular}
  \end{adjustbox}
\end{table*}

\begin{table*}
  \caption{Uni-QSAR ablation study on molecular property prediction classification tasks}
  \label{table:classification_ab}
  \centering
   \addtolength{\tabcolsep}{-2pt}
   \begin{adjustbox}{max width=\linewidth}
  \begin{tabular}{l|llllllll|lllll}
    \toprule
    \multicolumn{1}{c}{} &\multicolumn{11}{c}{Classification (Higher is better  $\uparrow$)}   &\multicolumn{1}{c}{}  \\
    \multicolumn{1}{c}{}  &\multicolumn{8}{c}{AUROC}  &\multicolumn{4}{c}{AUPRC}                 \\                  
    \midrule
    Datasets      & HIA  & Pgp  & Bioav  & BBB  & \thead{CYP3A4 \\ Substrate}  & hERG  & Ames  & DILI & \thead{CYP2C9 \\ Inhibition} & \thead{CYP2D6 \\ Inhibition} & \thead{CYP3A4 \\ Inhibition} & \thead{CYP2C9 \\ Substrate} & \thead{CYP2D6 \\ Substrate} \\
    \midrule


    Uni-QSAR w/o unimol        & {0.983}     & {0.922}       & {0.723}    & {0.917}    & {0.645}      & {0.855}     & {0.858}  & {0.912}  & {0.770} & {0.691} & {0.860} & {0.397}  & {0.713}\\

    Uni-QSAR w/o stacking        & {0.988}     & \textbf{0.934}       & {0.691}    & {0.913}    & \textbf{0.660}      & {0.841}     & {0.869}  & {0.926}  & {0.796} & {0.718} & {0.879} & {0.40}  & {0.720} \\
    \midrule
    Uni-QSAR        & \textbf{0.992}     & \textbf{0.934}       & \textbf{0.732}    & \textbf{0.925}    & {0.645}      & \textbf{0.856}     & \textbf{0.876}  & \textbf{0.942}  & \textbf{0.801} & \textbf{0.743} & \textbf{0.888} & \textbf{0.454} & \textbf{0.721} \\
    \bottomrule
  \end{tabular}
  \end{adjustbox}
\end{table*}

\graphicspath{{}}
\begin{figure}[!]
    \centering
    \includegraphics[width=3.2in]{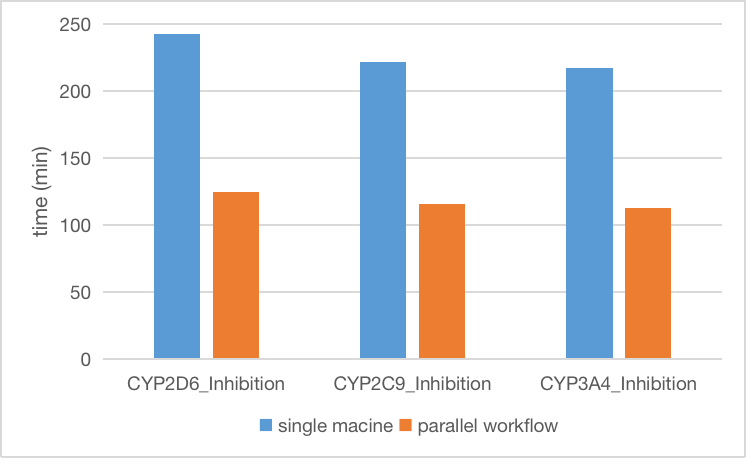}
    \caption{Uni-QSAR workflow Time efficiency. The CYP2C9 Inhibition, CYP2D6 Inhibition, and CYP3A4 Inhibition were chosen for benchmarking. All parallel workflows were running on Borihum with on-demand resource allocation}
    \label{fig:workflow_time_cost}
\end{figure}

\subsection{Time efficiency}
Uni-QSAR utilizes a combination of machine learning (ML) and deep learning (DL) models to achieve good performance. Training for these models can be time-consuming, so Uni-QSAR developed a workflow that can dispatch each model training job to different machines. What's more, compared with training on a single machine with GPU sequentially, when training the traditional ML model which runs on the CPU, the idle GPU leads to a significant waste of resources. Uni-QSAR assigns each model training job with the specific and appropriate machine configuration, thus GPU will not be wasted when the job runs on CPUs due to the on-demand resource allocation. To demonstrate the efficiency of the Uni-QSAR workflow, we compared the time cost between training on a single machine and on a parallel workflow. 
 The machine configuration consisted of 8 CPUs with 32GB memory and a V100 GPU, the parallel workflow version only requests V100 GPU on the DL model job.
 From Fig.\ref{fig:workflow_time_cost} shows that the Uni-QSAR workflow achieved 2x speedup when training on Borihum\footnote{https://bohrium.dp.tech/}, with notable improvement in real task.


\section{Conclusion}
In this paper, we proposed the first QSAR framework based on large-scale pretrained MRL called Uni-QSAR, and it is the first framework to combine 1D tokens, 2D topology graphs, and 3D conformers together in QSAR tools. The experimental results show that Uni-QSAR achieves state-of-the-art performance in 21 out of 22 tasks of the Therapeutic Data Commons (TDC) benchmark. 
The reason behind our framework to achieve the above effects through empirical evidence is that we use MRL to learn molecular representation information. 
It is worth noting that we have proved that the 3D structure of molecules plays a big role through ablation experiments. 
In addition, as an Auto ML and user-friendly tool, Uni-QSAR employs various strategies such as auto stacking and target normalization, allowing it to achieve good results on multiple datasets with a single set of parameters. 
At the same time, by using parallel workflow in Borihum, we can achieve a substantial utilization of resources, which improves the efficiency of the overall framework. 
What's more worth mentioning is that in order to illustrate the effectiveness and generalization of our framework, we used our framework in real CNS application scenarios, and we found that Uni-QSAR can achieve better generalization capabilities. \looseness=-1

\section{Future Work}
At present, our framework has provided commercialization and privatization support solutions, and we will adapt more property prediction scenarios under our framework to bring convenience to communities in many different fields, including small molecules, large molecules RNA, materials, and chemicals, etc. In order to improve the final effect of the entire framework, we will provide a more flexible hyper-parameter search interface, and we will optimize the DAG concurrency efficiency of the overall workflow. In addition to applying it to more attribute prediction scenarios, we will continue to optimize the interface to adapt to more challenging AI for science tasks, such as high throughput virtual screening.
\section{ACKNOWLEDGMENTS}
We thank Zhibo Chen, Xinzijian Liu, Weifeng Luo, Sikai Yao from the MLOPS team and many colleagues in DP Technology for their great help in this project.


\printbibliography


\clearpage
\appendix
\section{Appendix}
\paragraph{ADMET Benchmark Group}  ADMET Benchmark Group consists of 18 ADME and 4 Tox tasks, the Table \ref{table:datasets} shows the detailed description of 22 datasets.

\begin{table}[h]
  \caption{Descriptions of ADMET Benchmark Group datasets}
  \label{table:datasets}
  \centering
   \small
   \begin{adjustbox}{max width=\linewidth}
  \begin{tabular}{l|l|l|l}
    \toprule
    Dataset &Data size &Dataset Type &Task Type  \\ 
    \midrule
    Caco2  & 906     & Absorption  & Regression  \\
    HIA  & 578    & Absorption  & classification  \\
    Bioav  & 640    & Absorption  & classification  \\
    Pgp   & 1212   & Absorption  & classification  \\
    Lipo  & 4200    & Absorption  & Regression  \\
    AqSol  & 9982    & Absorption  & Regression  \\
    BBB   & 1975   & Distribution  & classification  \\
    PPBR  & 1797    & Distribution  & Regression  \\
    VDss  & 1130    & Distribution  & Regression  \\
    CYP2C9 Inhibition  & 12092    & Metabolism  & classification  \\
    CYP2D6 Inhibition  & 13130    & Metabolism  & classification  \\
    CYP3A4 Inhibition  & 12328    & Metabolism  & classification  \\
    CYP2C9 Substrate   & 666   & Metabolism  & classification  \\
    CYP2D6 Substrate   & 664   & Metabolism  & classification  \\
    CYP3A4 Substrate   & 667   & Metabolism  & classification  \\
    Half Life  & 667    & Excretion  & Regression  \\
    CL-Heap   & 1020   & Excretion  & Regression  \\
    Cl-Micro  & 1102   & Excretion  & Regression  \\
    LD50  & 7385    & Toxicity  & Regression  \\
    hERG  & 648    & Toxicity  & classification  \\
    Ames  & 7255    & Toxicity  & classification  \\
    DILI  & 475    & Toxicity  & classification  \\
    \bottomrule
  \end{tabular}
  \end{adjustbox}
\end{table}

\paragraph{Uni-QSAR Model Hyper-parameters} Table \ref{table:model_param} contains a list of ML and DL model architectures and associated hyper-parameters for producing the results in Table \ref{table:regression} and \ref{table:classification}. Note that for the linear model, we use logistic regression and ridge regression for the classification and regression tasks respectively. For the tree-based model, extraTree regressor \cite{scikit-learn} and LGBM regressor \cite{ke2017lightgbm} were used for the regression task while extraTree classifier \cite{scikit-learn} and LGBM classifier \cite{ke2017lightgbm} were used for the classification task.

\begin{table}[b]
  \caption{The parameter specification of the ensemble models}
  \label{table:model_param}
  \centering
   \small
   \begin{adjustbox}{max width=\linewidth}
  \begin{tabular}{l|l}
    \toprule
    Models & Model Hyper-parameters \\ 
    \midrule
    
    Linear Model \cite{scikit-learn} &  \thead{Inverse of regularization strength: 1.0} \\
    

    ExtraTree \cite{scikit-learn} &  \thead{ estimators: 100 \\ max depth: 40 \\ criterion: gini } \\
    
    LightGBM \cite{ke2017lightgbm}  & \thead{ estimators: 1000 \\ number of leaves: 31 \\ learning rate: 0.03 \\ subsample: 0.55 \\ colsample bytree: 0.55 } \\

    BERTModel \cite{intro8} &  \thead{ warmup\_ration: 0.03 \\ clip norm of the gradients: 5.0 \\ learning rate: 1e-4 \\ optimizer: Adam \\ scheduler: linear schedule with warmup } \\

    GNNModel \cite{hignn} & \thead{ warmup\_ration: 0.03 \\ clip norm of the gradients: 5.0 \\ learning rate: 1e-4 \\ optimizer: Adam \\ scheduler: linear schedule with warmup } \\

    Uni-Mol \cite{intro18} & \thead{ warmup\_ration: 0.03 \\ clip norm of the gradients: 5.0 \\ learning rate: 1e-4 \\ optimizer: Adam \\ scheduler: linear schedule with warmup } \\
    
    \bottomrule
  \end{tabular}
  \end{adjustbox}
\end{table}

\graphicspath{{}}
\begin{figure*}[h]
    \centering
    \includegraphics[width=14cm,height=8cm]{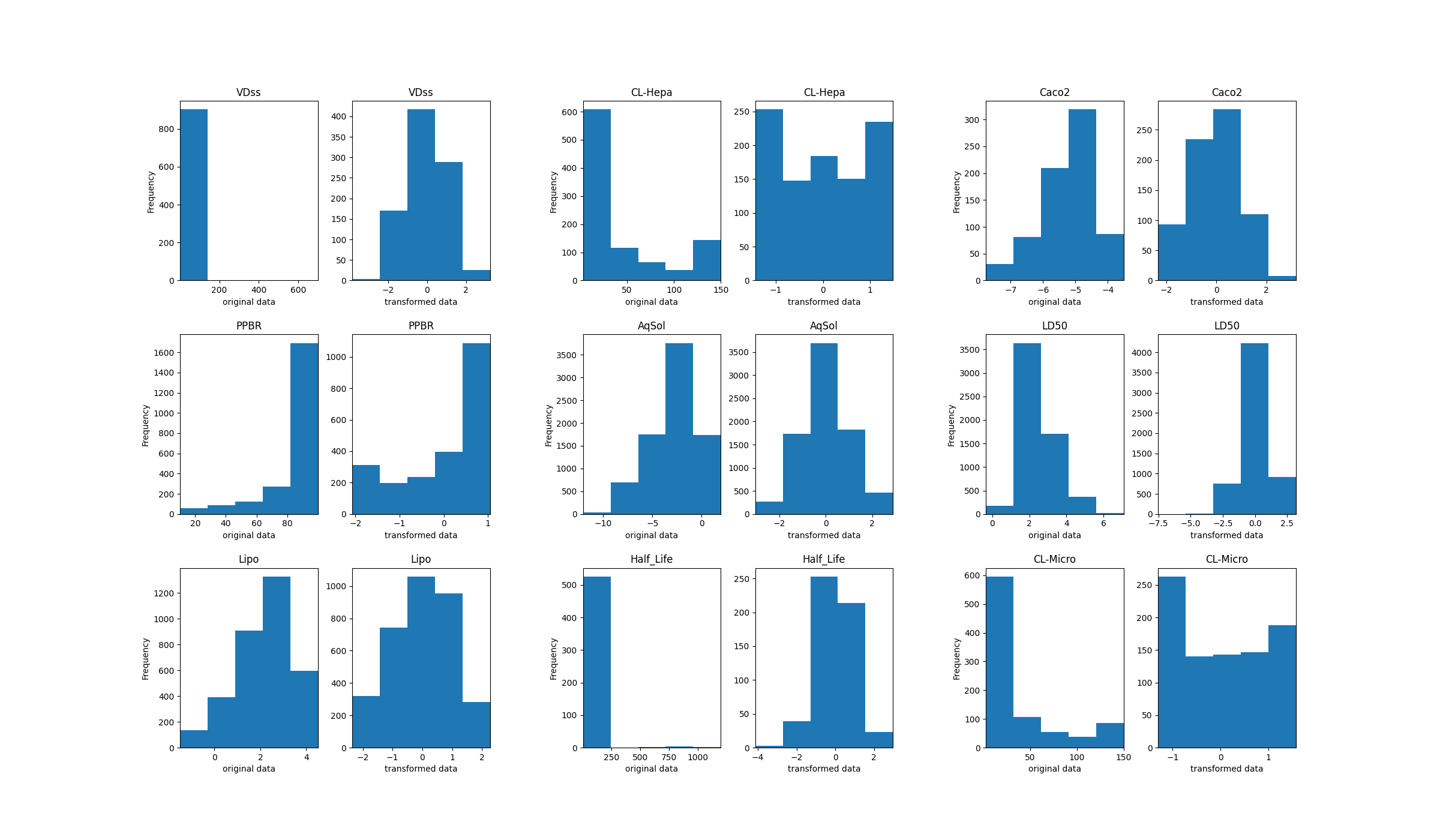}
    \vspace{-15pt}
    \caption{The Distribution of Regression's Task Data}
    \label{fig:distibution}
\end{figure*}

\paragraph{The Distribution of Regression Task Data} Fig. \ref{fig:distibution} shows the target data distribution and transformed target data distribution of all regression tasks. The \textit{VDss} and \textit{Half Life} task's data distribution is highly skewed before the transformation. The ablation study in Table \ref{table:regression_ab} shows that the performance improved
significant after performing normalization description in Alg. \ref{alg:targetnormal}.

\end{document}